\documentclass[sigconf]{acmart}
\usepackage{booktabs} 
\usepackage{algorithm}
\usepackage{algpseudocode}
\usepackage{wrapfig}
\usepackage{natbib}
\setcitestyle{numbers}

\AtBeginDocument{%
  }

\copyrightyear{2022}
\acmYear{2022}
\setcopyright{acmcopyright}\acmConference[CIKM '22]{Proceedings of the 31st ACM International Conference on Information and Knowledge Management}{October 17--21, 2022}{Atlanta, GA, USA}
\acmBooktitle{Proceedings of the 31st ACM International Conference on Information and Knowledge Management (CIKM '22), October 17--21, 2022, Atlanta, GA, USA}
\acmPrice{15.00}
\acmDOI{10.1145/3511808.3557148}
\acmISBN{978-1-4503-9236-5/22/10}

\begin{document}

\title{A Practical Distributed ADMM Solver for Billion-Scale Generalized Assignment Problems}



\author{Jun Zhou}
\authornote{Both authors contributed equally to this research.}
\affiliation{%
  \institution{$^1$ College of Computer Science and Technology \\ $^2$ Ant Group}
  \city{Hangzhou}
  \country{China}
}
\email{jun.zhoujun@antgroup.com}

\author{Feng Qi}
\authornotemark[1]
\affiliation{%
  \institution{Ant Group}
  \city{Sunnyvale}
  \country{USA}
}
\email{feng.qi@antgroup.com}

\author{Zhigang Hua}
\affiliation{%
  \institution{Ant Group}
  \city{Sunnyvale}
  \country{USA}
}
\email{z.hua@antgroup.com}

\author{Daohong Jian}
\affiliation{%
  \institution{Ant Group}
  \city{Beijing}
  \country{China}
}
\email{daohong.jdh@antgroup.com}

\author{Ziqi Liu}
\affiliation{%
  \institution{Ant Group}
  \city{Hangzhou}
  \country{China}
}
\email{ziqiliu@antgroup.com}

\author{Hua Wu}
\affiliation{%
  \institution{Ant Group}
  \city{Beijing}
  \country{China}
}
\email{wuhua.wh@antgroup.com}

\author{Xingwen Zhang}
\affiliation{%
  \institution{Ant Group}
  \city{Sunnyvale}
  \country{USA}
}
\email{xingwen.zhang@antgroup.com}

\author{Shuang Yang}
\authornote{denotes the corresponding author.}
\affiliation{%
  \institution{Ant Group}
  \city{Sunnyvale}
  \country{USA}
}
\email{shuang.yang@antgroup.com}

\renewcommand{\shortauthors}{Jun Zhou et al.}

\begin{abstract}
Assigning items to owners is a common problem found in various real-world applications, for example, audience-channel matching in marketing campaigns, borrower-lender matching in loan management, and shopper-merchant matching in e-commerce. Given an objective and multiple constraints, an assignment problem can be formulated as a constrained optimization problem. Such assignment problems are usually NP-hard \cite{martello1990knapsack}, so when the number of items or the number of owners is large, solving for exact solutions becomes challenging. In this paper, we are interested in solving constrained assignment problems with hundreds of millions of items. Thus, with just tens of owners, the number of decision variables is at billion-scale. This scale is usually seen in the internet industry, which makes decisions for large groups of users.
We relax the possible integer constraint, and formulate a general optimization problem that covers commonly seen assignment problems. Its objective function is convex. Its constraints are either linear, or convex and separable by items. We study to solve our generalized assignment problems in the Bregman Alternating Direction Method of Multipliers (BADMM) framework where we exploit Bregman divergence to transform the Augmented Lagrangian into a separable form, and solve many subproblems in parallel. The entire solution can thus be implemented using a MapReduce-style distributed computation framework. We present experiment results on
both synthetic and real-world datasets to verify its accuracy and scalability.
\end{abstract}

\ccsdesc[500]{Mathematics of computing~Integer programming}
\ccsdesc[300]{Theory of computation~MapReduce algorithms}

\keywords{Assignment Problem; ADMM; Large-Scale Optimization; Distributed Algorithms}

\maketitle

\section{Introduction}
Many real-world scenarios can be abstracted as assigning "items" to "owners" (or "bins"). To name a few, assigning customers to channels in marketing campaigns \cite{liu2019graph,yu2021joint}, assigning borrowers to lenders in loan management~\cite{cheng2020contagious}, and assigning shoppers to merchants in e-commerce \cite{wang2016selecting} are assignment examples that are designed to maximize or minimize some metrics while simultaneously satisfying requirements on other metrics. 
The assignment problems in internet industry could have
multiple variants, and the scale of decision variables
could be in billions. Existing literature either study
particular forms of assignment problems or cannot solve
problems in billion scale. In this paper, we formulate a more general assignment problem, and propose a practical
distributed solver to decompose the Augmented Lagrangian of the original large-scale problem into small pieces via Bregman divergence under BADMM framework~\cite{wang2013bregman}. We show the quality of the solution is promising in real-world applications.
\subsection{Problem Formulation} \label{ProblemFormulation}
 To assign $I$ items to $J$ owners, we use $X \in \{0,1\}^{I\times J}$ or $X \in [0,1]^{I\times J}$ to represent the decision variables. In the former case, the element in row $i$ and column $j$ $x_{i,j} \in \{0,1\}$ is 1 if item $i$ is assigned to owner $j$, and 0 if not. In the latter case, $x_{i,j} \in [0,1]$ indicates that a portion of item $i$ is assigned to owner $j$. We also use $x_i \in \{0,1\}^I$ or $x_i \in [0,1]^I$ to represent the $i$-th row of $X$.

We study the following generalized constrained assignment problem: (\ref{Objective})--(\ref{IntegerConstraint}),
\begin{flalign}
\min_{X} & \displaystyle \quad f(X) \label{Objective} \\
s.t. & \displaystyle \quad X^T u_m \leq b_m, \quad \forall m \in [M] \label{Inequality} \\
& \displaystyle \quad X^T v_n = c_{n}, \quad \forall n \in [N]  \label{Equality} \\
& \displaystyle \quad h_i(x_i) \leq 0, \quad \forall i \in [I] \label{LocalConstraint}\\
& \displaystyle \quad x_{i,j} \in \{0,1\} \quad \text{or} \quad 0 \leq x_{i,j} \leq 1 \quad \forall i \in [I], \forall j \in [J]. \label{IntegerConstraint}
\end{flalign}
We require that when $X$ is continuous on $[0, 1]^{I\times J}$, the objective function $f(X)$ is convex. The first set of constraints (\ref{Inequality}) are $J\times M$ inequalities, with coefficients $u_m \in \mathbb{R}^{I}$ and $b_m \in \mathbb{R}^J$. They allow each of the $J$ owners to have one linear inequality constraint on each of a set of $M$ item features. The second set of constraints (\ref{Equality}) are $J\times N$ equations, with coefficients $v_n \in \mathbb{R}^{I}$ and $c_n \in \mathbb{R}^J$. They allow each of the $J$ owners to have one linear equality constraint on each of another set of $N$ item features. The third set of constraints (\ref{LocalConstraint}) are $I$ inequalities each on one item $x_i$. We allow each item to have its unique $h_i(x_i)$. The most commonly used such constraint is $h_i(x_i)=\sum_j x_{i,j} - 1 \leq 0$. We distinguish between inequality (\ref{Inequality}) and equality (\ref{Equality}) constraints because the former needs an extra treatment in our algorithm. For item-level constraints (\ref{LocalConstraint}), we only consider inequality without loss of generality. Constraint (\ref{IntegerConstraint}) either requires each $x_{i,j}$ to be a binary decision, or allows a portion of each item $i$ to be allocated to an owner $j$.

\subsection{Our Work} \label{Background}
There exist three main challenges in solving the above assignment problems:
\begin{itemize}
    \item The scale of billions of decision variables is unprecedented. We implemented distributed algorithm on top of the MapReduce framework, and verified its scalability and convergence in experiments and production.
    \item The assignment problems are NP-hard integer programming (IP) problems. We solve the linear programming (LP) relaxation of the original IP problems and project the solutions back to integers. We empirically verify this practical approach yields results that are close to optimal and acceptable.
    \item The complexity of solving the assignment problems is further increased when the objective function is nonlinear and non-separable. Though the BADMM is a general framework to deal with quadratic penalties, the use of Bregman divergence~\cite{wang2013bregman} to derive decomposable forms for our general assignment problems is still non-trivial.
\end{itemize}
As far as we know, this has been one of the first works to solve the assignment problems in billion-scale. The experiments have shown promising results in terms of both accuracy and convergence. 

\section{Related Work}\label{Related}
\subsection{Generalized Assignment Problem (GAP)}
GAP~\cite{martello1990knapsack} is a well studied problem, and can be viewed as a generalization of knapsack problem or bin-packing problem. It solves for the minimum cost when assigning $I$ items to $J$ owners. Each owner has a capacity $r_j$. Each item is assigned to one owner. Its mathematical formulation is

\begin{flalign}
\min_{X} & \displaystyle \quad \sum_i\sum_j p_{ij}x_{ij} \label{ObjectiveGAP} \\
s.t. & \displaystyle \quad \sum_i q_{ij} x_{ij} \leq r_j, \quad \forall j \in [J] \label{InequalityGAP} \\
& \displaystyle \quad \sum_j x_{ij} = 1, \quad \forall i \in [I] \label{LocalConstraintGAP}\\
& \displaystyle \quad x_{i,j} \in \{0,1\} \quad \forall i \in [I], \forall j \in [J]. \label{IntegerConstraintGAP}
\end{flalign}
In (\ref{ObjectiveGAP})-(\ref{IntegerConstraintGAP}), $p_{ij}\in \mathbb{R}_{\geq 0}$ is the cost of assigning item $i$ to owner $j$, and $q_{ij}\in \mathbb{R}_{\geq 0}$ is the consumption of capacity if assigning item $i$ to owner $j$.

GAP is a linear IP problem, and is NP-hard. To solve for exact solutions, branch-and-bound and branch-and-price methods can be used \cite{savelsbergh1997branch}.

When exact solution is not necessary, one class of algorithms solves GAP by relaxing constraints. \citet{ross1975branch} proposed an algorithm that first relaxes (\ref{InequalityGAP}) to get a lower bound result, and then solves $J$ knapsack problems. \citet{martello1981algorithm} relaxes (\ref{LocalConstraintGAP}), and solves $J$ knapsack problems, while a penalty is applied to alleviate constraint violation. \citet{jornsten1986new} proposed a Lagrangian relaxation algorithm, where the optimal Lagrangian multipliers are obtained by a subgradient method. \citet{fisher1986multiplier} developed a multiplier adjusting method to determine good Lagrangian multipliers. Linear programming relaxation (LP-relaxation) can also be applied by replacing (\ref{IntegerConstraintGAP}) with $x_{ij} \geq 0$. In "practical" problems where $I >> J$, this relaxation leds to solutions that are within $0.1\%$ of their LP lower bounds, while in settings where $\frac{J}{I}$ equals 3 or 4, reassigning the fractional jobs can turn out to be infeasible \cite{cattrysse1992survey}.

Another class of algorithms solve GAP by heuristic optimization, for example, genetic algorithms \cite{chu1997genetic}\cite{feltl2004improved}, simulated annealing and tabu search \cite{osman1995heuristics}, heuristic search and ant system \cite{lourencco1998adaptive}.

Variants of GAP are also studied. In Multi-Resource GAP, each owner has multiple resource constraints \cite{gavish1991algorithms}. In Multilevel GAP, a third dimension "level" is added to each item and owner combination, thus each assignment $x_{ij}$ operating at  level $k$ is represented as $x_{ijk}$ \cite{avella2013branch,ceselli2006branch}. Standard GAP considers resource upper bounds, whereas \cite{krumke2013generalized} considered lower bounds. \cite{xiong2017new} discussed the scenario when multiple items from each of $K$ categories are required by each owner, and each item belongs to a subset of the $K$ categories.

Our formulation (\ref{Objective})-(\ref{IntegerConstraint}) is different from GAP. First, we allow nonlinear objective (\ref{Objective}) and nonlinear per-item constraints (\ref{LocalConstraint}). Second, we allow each item feature $u_m$ and $v_n$ to interact with each owner $j$ in (\ref{Inequality}) and (\ref{Equality}). GAP is actually a special case of (\ref{Objective})-(\ref{IntegerConstraint}). To see this:
\begin{itemize}
    \item Notice (\ref{ObjectiveGAP}) and (\ref{LocalConstraintGAP}) are specified (\ref{Objective}) and (\ref{LocalConstraint}), respectively.
    \item To convert (\ref{InequalityGAP}) to (\ref{Inequality}), set $M=J$ to get $J$x$J$ inequality constraints. When $m \neq j$, set the $j$-th element in $b_{m}$ to infinity.
    \item To remove (\ref{Equality}), set all elements of $c_n$ to infinity.
\end{itemize}

In fact, the constraint structure of GAP, Multi-Resource GAP, and our formulation are all just "$Ax\leq b$" with different block matrix patterns, when we use a vector $x \in \mathbb{R}^{IJ \times 1}$ to represent decision variables. ADMM discussed in Section \ref{ADMMSS} can solve all these systems.

\subsection{ADMM}\label{ADMMSS}
ADMM \cite{boyd2011distributed} is an optimization framework for constrained optimization. It uses the Augmented Lagrangian function as objective. For an equality-constrained optimization, the objective adds quadratic penalties on constraint violation to the Lagrangian function. ADMM can be extended to work with inequality constraints by adding slack variables.

An ADMM method iteratively solves primal and dual variables until a stopping criterion is met. To solve for primal variables, it separates variables into two or more blocks, and solves them sequentially with Gauss-Seidel routine \cite{bertsekas1989parallel}. Due to the quadratic penalty in the Augmented Lagrangian, solving each block can be viewed as a proximal problem \cite{parikh2014proximal}. To solve for dual variables, ADMM uses a subgradient method with a pre-determined learning rate. It was shown that when the objective function of the primal problem is a summation of convex functions, ADMM enjoys a linear convergence rate on primal and dual variables \cite{hong2017linear}, and $O(1/T)$ convergence rate on objective \cite{he20121}, where $T$ is the number of iterations. ADMM can also be used in non-convex optimization, such as mixed-integer quadratic programming, though there is a lack of convergence guarantee, and hence the usage needs to be justified by experiments \cite{takapoui2020simple}.

When the Augmented Lagrangian is difficult to solve, each block of variables may be constructed easier to solve. For a large-scale optimization problem, if we can make each block separable on decision variables, we can solve sub-problems in parallel with Jacobi routine \cite{bertsekas1989parallel}. The non-separability of Augmented Lagrangian may be due to either the quadratic penalty or the primal objective. Proximal Jacobi ADMM \cite{deng2016global,deng2017parallel} adds to the Augmented Lagrangian a Mahalanobis distance between the primal variables and their values in the previous iteration. By choosing the Mahalanobis distance carefully, any quadratic penalties due to the general form of equality constraints "$Ax=b$" can be reduced to separable terms. Generalized Bregman ADMM (BADMM) \cite{wang2013bregman} generalizes the quadratic penalty to a Bregman divergence. The proximal term between primal variables and their values in the previous iteration is modeled as another Bregman divergence. By constructing these Bregman divergences carefully, any non-separable terms in the Augmented Lagrangian can be reduced to separable terms. BADMM enjoys $O(1/T)$ convergence on objective, when the objective is convex, and the penalty Bregman divergence is defined on a $\alpha$-strongly convex function, where $\alpha > 0$. Linearized ADMM \cite{liu2019linearized} uses first-order approximation and second order proximal terms to approximate the Augmented Lagrangian in ADMM iterations, and hence making the problem separable.

In this paper, since the use of Bregman divergence could help us effectively exploit the structure of problems, we follow the framework of BADMM~\cite{wang2013bregman} and use Bregman divergence to derive decomposable sub-problems for our generalized assignment problems
with even non-separable objectives so that it can be solved in parallel.


\section{Our Approach} \label{Approach}
To handle large scale items ($I$) and owners ($J$), we solve (\ref{Objective}) - (\ref{IntegerConstraint}) in the BADMM framework. This framework updates primal and dual variables iteratively, as ADMM does, and decomposes the objective function $f(X)$ into a separable form via replacing non-separable terms with Bregman divergence terms. Therefore the primal update step in ADMM can be divided into many subproblems and solved parallelly. 
To discuss the detailed routine, we first form the Augmented Lagrangian. We rewrite (\ref{Objective})--(\ref{IntegerConstraint}) by moving (\ref{LocalConstraint}) and (\ref{IntegerConstraint}) into the objective. We also convert (\ref{Inequality}) into equality constraints by introducing additional positive decision variables $\xi_m \in \mathbb{R}_{\geq0}^M$:
\begin{flalign}
\min_{X,\xi} & \displaystyle \quad f(X) + \sum_i\Phi(h_i(x_i)) + \sum_{m,j}\Phi(-\xi_{m,j}) + \sum_{i,j}\Psi(x_{i,j}) \label{ObjectiveAL} \\
s.t. & \displaystyle \quad X^T u_m + \xi_m = b_m, \quad \forall m \in [M] \label{InequalityAL} \\
& \displaystyle \quad X^T v_n = c_{n}, \quad \forall n \in [N].  \label{EqualityAL}
\end{flalign}
In (\ref{ObjectiveAL}), $\Phi$ and $\Psi$ are defined as below.
\begin{flalign}
& \Phi(y) =
    \begin{cases}
        0,& y \leq 0 \\
        +\infty,& y > 0
    \end{cases} \\
&\Psi(y) =
    \begin{cases}
        0,& \text{$y$ satisfies (\ref{IntegerConstraint}}) \\
        +\infty,& \text{otherwise}.
    \end{cases}
\label{PhiPsi}
\end{flalign}
Therefore (\ref{ObjectiveAL}) - (\ref{EqualityAL}) are equivalent to (\ref{Objective})--(\ref{IntegerConstraint}). The Augmented Lagrangian based on (\ref{ObjectiveAL}) - (\ref{EqualityAL}) is as below.

\begin{equation}
\begin{split}
L(X, \xi, \lambda, \mu) = & \displaystyle \quad f(X) + \sum_i\Phi(h_i(x_i)) + \sum_{m,j}\Phi(-\xi_{m,j}) + \sum_{i,j}\Psi(x_{i,j}) \\
& + \frac{\rho}{2}\sum_m \Vert X^T u_m + \xi_m - b_m + \frac{\lambda_m}{\rho} \Vert_2^2 \\
& + \frac{\rho}{2}\sum_n \Vert X^T v_n - c_n + \frac{\mu_n}{\rho} \Vert_2^2. \label{AugLag} \\
\end{split}
\end{equation}
Here we use $\lambda_m \in \mathbb{R}^J$ to represent the Lagrangian multipliers of (\ref{InequalityAL}), and $\mu_n \in \mathbb{R}^J$ to represent the Lagrangian multipliers of (\ref{EqualityAL}). $\rho$ is a predetermined constant for the quadratic penalty terms. We also use $\xi$ to represent the collection $\xi_{1...M}$, $\lambda$ to represent the collection $\lambda_{1...M}$, and $\mu$ to represent the collection $\mu_{1...N}$. 

Using ADMM, we can solve (\ref{AugLag}) by iteratively run the following four steps
described in Algorithm~\ref{alg:ADMMAlgo}. We use superscript to label the iteration in which a variable is determined.

\begin{algorithm} 
\caption {The ADMM routine for solving (\ref{ObjectiveAL}) - (\ref{EqualityAL}) by (\ref{AugLag})}
\label{alg:ADMMAlgo}
\begin{algorithmic}[1]
\State \textbf{Input:} Iteration t = 0, and initialize $X$, $\xi$, $\lambda$, $\mu$ with $0$.
\State \textbf{Output:} X
\While{Not converged}
    \State $X^{(t+1)} = \arg\min_X L(X, \xi^{(t)}, \lambda^{(t)}, \mu^{(t)})$
    \State $\xi^{(t+1)} = \arg\min_{\xi} L(X^{(t+1)}, \xi, \lambda^{(t)}, \mu^{(t)})$
    \State $\lambda_m^{(t+1)} = \lambda_m^{(t)} + \rho((X^{(t+1)})^T u_m + \xi_m^{(t+1)} - b_m)\quad \forall m$
    \State $\mu_n^{(t+1)} = \mu_n^{(t)} + \rho((X^{(t+1)})^T v_n - c_n)\quad \forall n$
    \State $t = t + 1$
\EndWhile
\end{algorithmic}
\end{algorithm}

\subsection{ADMM Routine}\label{ADMMMethod}
Here we introduce the following steps in each ADMM iteration as per line 4-7 in Algorithm~\ref{alg:ADMMAlgo}.

\textbf{Step 1: Update $X$}. We first update the decision variables $X$ as follows:
\begin{equation}
\begin{split}
X^{(t+1)} = \arg\min_{X} & \quad f(X) + \sum_i\Phi(h_i(x_i)) + \sum_{i,j}\Psi(x_{i,j}) \\
& + \frac{\rho}{2}\sum_m \Vert X^T u_m + \xi_m^{(t)} - b_m + \frac{\lambda_m}{\rho} \Vert_2^2 \\
& + \frac{\rho}{2}\sum_n \Vert X^T v_n - c_n + \frac{\mu_n}{\rho} \Vert_2^2. \label{ADMMS1} \\
\end{split}
\end{equation}
Since $\xi_m^{(t)}$ is known at this step, it can be absorbed by $b_m$. The last two quadratic terms thus have the same pattern, and can be merged: 
\begin{equation}
\begin{split}
X^{(t+1)} = \arg\min_{X} & \quad f(X) + \sum_i\Phi(h_i(x_i)) + \sum_{i,j}\Psi(x_{i,j}) \\
& + \frac{\rho}{2}\sum_s \Vert X^T w_s - d_s + \frac{\nu_s}{\rho} \Vert_2^2. \label{ADMMS1b} \\
\end{split}
\end{equation}
In (\ref{ADMMS1b}), we used index $s$ to combine indices $m$ and $n$, $w_s$ to replace $u_m$ or $v_n$, $d_s$ to replace $b_m - \xi^{(t)}_m$ or $c_n$, and $\nu_s$ to replace $\lambda_m$ or $\mu_n$. Due to the inseparable form, we discuss the solving of this problem in parallel in Section~\ref{sec:badmm}.


\textbf{Step 2: Update $\xi$}.
To update $\xi$, we drop constant terms from (\ref{AugLag}).
\begin{equation}
\begin{split}
\xi^{(t+1)}= &\quad \arg\min_{\xi} \sum_{m,j}\Phi(-\xi_{m,j})\\ &\quad + \frac{\rho}{2}\sum_m \Vert (X^{(t+1)})^T u_m + \xi_m - b_m + \frac{\lambda_m^{(t)}}{\rho} \Vert_2^2. \label{ADMMs2} \\
\end{split}
\end{equation}
This problem is equivalent to solving $M \times J$ 1-variable constrained quadratic programming problem:
\begin{equation}
\begin{split}
\min_{\xi_{m,j}} &\quad (\xi_{m,j} - l_{m,j})^2 \\
s.t. & \quad \xi_{m,j} \geq 0.
\end{split}
\end{equation}
Here $l_m = -(X^{(t+1)})^T u_m + b_m - \frac{\lambda_m^{(t)}}{\rho}$, and $l_{m,j}$ is the j-th element of $l_m$. The solution to the problem is
\begin{equation}\label{XIUpdate}
\begin{split}
\xi_{m,j} = \max(0, l_{m,j}).
\end{split}
\end{equation}

\textbf{Step 3-4: Update $\lambda$ and $\mu$}. 
The last two steps are to update the Lagrangian multipliers.
\begin{equation}
\begin{split}
\lambda_m^{(t+1)} = \lambda_m^{(t)} + \rho((X^{(t+1)})^T u_m + \xi_m^{(t+1)} - b_m) \\
\mu_n^{(t+1)} = \mu_n^{(t)} + \rho((X^{(t+1)})^T v_n - c_n).
\end{split}
\end{equation}
According to (\ref{XIUpdate}), if $\xi_m^{(t+1)}$ is larger than 0, $\lambda_m^{(t+1)}$ will be set to 0 in the $t+1$-th iteration.

\subsection{Decomposition via Bregman Divergence}\label{sec:badmm}
(\ref{ADMMS1b}) is still a difficult problem given that the scale of $X$ is large and that the exact formula of $f(X)$ is nonseparable by elements of $X$. We derive to convert (\ref{ADMMS1b}) into a separable function of $X$
under the BADMM framework. Basically, we introduce a Bregman divergence term into the Augmented Lagrangian.

\begin{equation}
\begin{split}
X^{(t+1)} = \arg\min_{X} & \quad f(X) + \sum_i\Phi(h_i(x_i)) + \sum_{i,j}\Psi(x_{i,j}) \\
& + \frac{\rho}{2}\sum_s \Vert X^T w_s - d_s + \frac{\nu_s}{\rho} \Vert_2^2 + B_{\Omega}(X,X^{(t)}) \label{ADMMS1c} \\
\end{split}
\end{equation}

The Bregman divergence $B_{\omega}(x,y)$ defined on a convex function $\omega(z) \in \mathbb{R}$ is $\omega(x)-\omega(y)-\omega'(y)^T(x-y)$. Our decision variable $X$ is a matrix in $\mathbb{R}^{I\times J}$. For convenience we slightly abuse the derivative symbol $\Omega'(X)$ to let it represent a $\mathbb{R}^{I\times J}$ matrix, the element on the $i$-th row and $j$-th column of which is the derivative of $\Omega(X)\in \mathbb{R}$ to $x_{i,j}$. We use $\circ$ to represent Hadamard product. We also use the operation $\mathbf{1}^TA\mathbf{1}$ to sum up all elements of a matrix $A \in \mathbb{R}^{I\times J}$. In this case the first $\mathbf{1}$ is a vector of ones in $\mathbb{R}^I$, and the second $\mathbf{1}$ is a vector of ones in $\mathbb{R}^J$. We use the same symbol for these two vectors for simplicity, since their dimensions are known once $A$'s dimension is known. The Bregman divergence term in (\ref{ADMMS1c}) can be expressed as:
\begin{equation}\label{BregmanDivergence}
B_{\Omega}(X,X^{(t)}) = \Omega(X) - \Omega(X^{(t)}) - \mathbf{1}^T\Omega'(X^{(t)}) \circ (X - X^{(t)})\mathbf{1} \\
\end{equation}
In our design, $\Omega$ is composed of two components $\Omega_1$ and $\Omega_2$.
\begin{flalign}
 & \quad \Omega = \Omega_1 + \Omega_2 \label{OmegaBoth} \\
\text{where} & \quad \Omega_1=g(X) - f(X) \label{Omega1} \\
 & \quad \Omega_2=\beta \operatorname{tr}(X^T X) - \frac{\rho}{2}\sum_s w_s^T X X^T w_s \label{Omega2}
\end{flalign}
The solution can be obtained as follows:
\begin{equation}
\begin{split}
X^{(t+1)} = \arg\min_{X} & \quad \sum_k g(X_k) - \mathbf{1^T}(g'(X^{(t)}) - f'(X^{(t)})) \circ X \mathbf{1}\\
& + \sum_i\Phi(h_i(x_i)) + \sum_{i,j}\Psi(x_{i,j}) \\
& + \beta \operatorname{tr}((X - X^{(t)})^T (X - X^{(t)})) \\
& +\rho\mathbf{1}^T(\sum_s w_s (X^{(t)T} w_s  - d_s+\frac{\nu_s}{\rho})^T)\circ X\mathbf{1}\label{ADMMS1z}
\end{split}
\end{equation}
Symbols $g'$ and $f'$ are used in a similar way as $\Omega'$, each element of both matrices represents the derivative with respect to the corresponding element of $X$. $\beta$ is chosen such that $\beta \geq \frac{\rho}{2} I (M+N)$. The function $g(X)$ is constructed such that it is separable on the first dimension of $X$, i.e. $g(X) = \sum_k g_k(X_k)$, where $\{k\}$ is a mutually exclusive and collectively exhaustive set of $X$ row index clusters. We assume such $g(X)$ can be found by choosing a convex $\Omega_1$. In Section \ref{ExperimentsSection} we provide concrete examples.

In (\ref{ADMMS1z}) each term is separable by $X$'s row index clusters $\{k\}$. We use subscript $k$ to indicate taking the $i\in k$ rows from a $\mathbb{R}^{I\times J}$ matrix and form a $\mathbb{R}^{\vert k\vert\times J}$ matrix. (\ref{ADMMS1z}) can be divided into $|k|$ independent optimization problems.

\begin{equation}
\begin{split}
\min_{X_k} & \quad g(X_k) + \mathbf{1^T}B_k \circ X_k \mathbf{1} + \beta \operatorname{tr}((X_k - X^{(t)}_k)^T (X_k - X^{(t)}_k)) \\
s.t. &\quad h_i(x_i) \leq 0 \quad \forall i \in k \\
&\quad x_{i,j} \in \{0,1\} \quad\text{or}\quad 0 \leq x_{i,j} \leq 1 \quad \forall i\in k,j \\
\text{where} &\quad B = -g'(X^{(t)}) + f'(X^{(t)}) +\rho\sum_s w_s (X^{(t)T} w_s  - d_s+\frac{\nu_s}{\rho})^T \label{ADMMS1y}
\end{split}
\end{equation}
We have moved the non-differentiable components $\Psi$ and $\Phi$ to constraints to illustrate that (\ref{ADMMS1y}) is a constrained convex optimization problem. In practice $X_k$ can usually be made much smaller than $X$ in size, and the $|k|$ problems of (\ref{ADMMS1y}) can be solved in parallel using existing convex optimization techniques.

\subsection{Acceleration For Quadratic Objective On Simplex Region} \label{AccelerateQP}

In practice, we found that $g(X)$ can often chosen to be a quadratic function that is separable over $x_{ij}$, i.e. $g(X) = \sum_i\sum_j a_{ij}x_{ij}^2$ where $a_{ij}\in R$. Two concrete examples are presented in Section \ref{ExperimentsSection}. The per-item constraint $h_i(x_i) \leq 0$ is usually $\sum_j x_{ij} \leq 1$. Thus each block $k$ in (\ref{ADMMS1y}) can be further separated into many optimization problems for each item $i$.

\begin{equation}
\begin{split}
\min_{x_{ij}} & \quad \sum_j a_{ij}x_{ij}^2 + B_{ij} x_{ij} + \beta (x_{ij} - x^{(t)}_{ij})^2 \\
s.t. &\quad \sum_j x_{ij} \leq 1 \quad \text{and}
\quad x_{i,j} \geq 0 \quad \forall j. \label{ADMMS1g}
\end{split}
\end{equation}
$B_{ij}$ is the element of B on the $i$-th row and $j$-th column. Transform (\ref{ADMMS1g}) to simplify its formula.

\begin{equation}
\begin{split}
\min_{x_{ij}} & \quad \sum_j \gamma_{ij}x_{ij}^2 + \eta_{ij} x_{ij} \\
s.t. &\quad \sum_j x_{ij} \leq 1 \quad \text{and} \quad x_{i,j} \geq 0 \quad \forall j \\
\text{where} & \quad \gamma_{ij} = a_{ij} + \beta, \quad \eta_{ij} = B_{ij}-2\beta x_{ij}^{(t)}.
\label{ADMMS1h}
\end{split}
\end{equation}
This sub-problem can be solved with $O(J\ln{J}$) time complexity. We formulate its partial Lagrangian
\begin{equation}
\begin{split}
\max_{\pi}\min_{x_{ij}} & \quad \sum_j \gamma_{ij}x_{ij}^2 + \eta_{ij} x_{ij} + \pi(\sum_j x_{ij} - 1) \\
s.t. &\quad x_{i,j} \geq 0 \quad \forall j, \quad \pi \geq 0.
\label{ADMMS1i}
\end{split}
\end{equation}
The solution to the inner minimization problem, considering the constraint on $x_{ij}$, is $x_{ij}=\max(0, -\frac{\eta_{ij}+\pi}{2\gamma_{ij}}), \quad \forall j$. Substituting $x_{ij}$ with this solution gives

\begin{equation}
\begin{split}
\max_{\pi} & \quad \sum_j  (\gamma_{ij}max(0, \frac{\eta_{ij}+\pi}{2\gamma_{ij}})^2-\frac{(\eta_{ij}+\pi)^2}{4\gamma_{ij}})-\pi\\
&\quad \pi \geq 0.
\label{ADMMS1j}
\end{split}
\end{equation}
This is a non-differentiable concave function. Assume we have thresholds $-\eta_{ij},\quad j=0,1..J-1$ sorted in ascending order, and let the sorted non-negative values be $\{-\eta_{ij'}\}$. When $\pi$ is on range $[\eta_{ij'}, \eta_{ij'+1}]$, (\ref{ADMMS1j}) is:

\begin{equation}
\begin{split}
\max_{\pi} & \quad \sum_{j\in J'}  -\frac{(\eta_{ij}+\pi)^2}{4\gamma_{ij}}-\pi\\
s.t &\quad -\eta_{ij'} \leq \pi \leq -\eta_{ij'+1} \\
\text{where} & \quad J' = \{j|\eta_{ij} \leq \eta_{ij'+1}\}.
\label{ADMMS1k}
\end{split}
\end{equation}
The solution to this $\pi$ range is
\begin{equation}
\begin{split}
& \pi_j^*=\max(-\eta_{j'}, \min(-\eta_{j'+1}, -\frac{\sum_{j\in J'}\frac{\eta_{ij}}{2\gamma_{ij}}+1}{\sum_{j\in J'}\frac{1}{2\gamma_{ij}}}))\\
&\quad \text{where} \quad J'=\{j|\eta_{ij} \leq \eta_{ij'+1}\}.
\label{ADMMS1l}
\end{split}
\end{equation}
The corresponding objective for $\pi_j^*$ is
\begin{equation}
\begin{split}
& -\sum_{j\in J'}\frac{1}{4\gamma_{ij}} \pi_j^{*2}-(\sum_{j \in J'}\frac{\eta_{ij}}{2\gamma_{ij}}+1)\pi_j^* - \sum_{j\in J'}\frac{\eta_{ij}^2}{4\gamma_{ij}}.
\label{ADMMS1m}
\end{split}
\end{equation}
Let $\pi^*$ be the $\pi_j^*$ with the maximum objective (\ref{ADMMS1m}), thus the corresponding solutions to $x_{ij}$ is $x_{ij}^*=\max(0, -\frac{\eta_{ij}+\pi^*}{2\gamma_{ij}})$).

Consider the entire procedure, an outer loop running a $O(\ln{J})$ binary search over the concave (\ref{ADMMS1j}) is used to search for a range of $\pi$. At each range of $\pi$, $\pi_j^*$ and its objective are evaluated with $O(J)$ time. It is possible to pre-compute $\pi_j^*$ and the three coefficients for its objective with $O(J)$ time, though an initial sort of $\eta_{ij}$ is required for this pre-computation, so the overall complexity is still $O(J\ln{J})$.

\subsection{Post-processing for Integer Solutions}
When integer solution is desired, we use post-process to convert continuous solutions to integers. As discussed in Section \ref{Related}, when the number of decision variables is small, post-process like rounding can lead to infeasiblity or suboptimality, but in large scale problems it empirically generates near-optimal solutions. Given local constraints like $\sum_j x_{ij}\leq 1, \forall i \in [I]$, our post-processing treats each $x_{ij}$ as the probability of assigning item $i$ to owner $j$, and the probability of not assigning $i$ to any $j$ is $1-\sum_j x_{ij}$. We sample a binary solution from such probability simplex for each item, to obtain integer solutions. 

\subsection{Distributed System Design}

\begin{figure}
\centering
\includegraphics[width=0.7\linewidth]{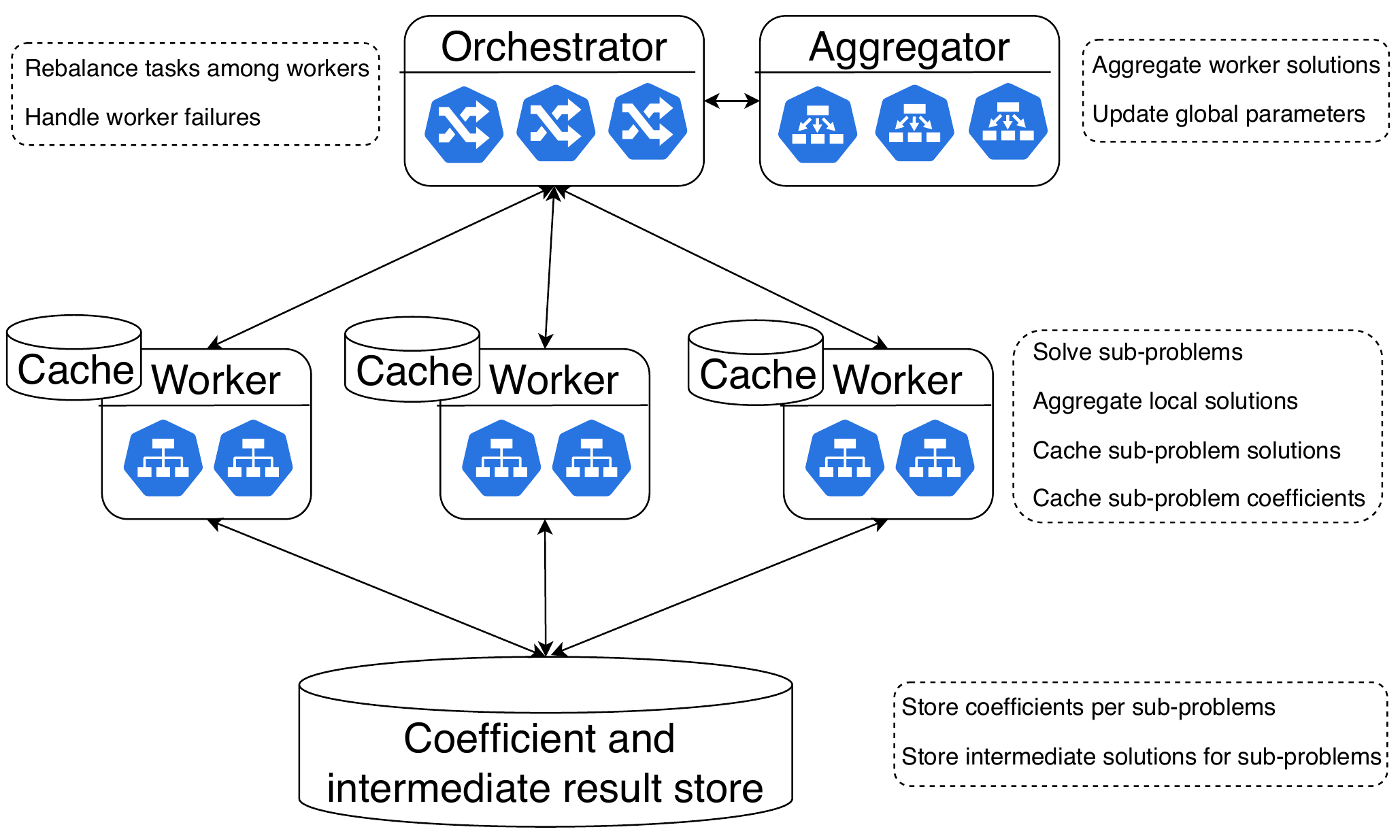}
\caption{The system of our distributed solver.}
\label{SystemGraph}
\end{figure}

Figure \ref{SystemGraph} shows the system construction of our distributed solver, which consists of an orchestrator, an aggregator, a farm of workers, and a permanent data store.

Coefficients of $g$, $u_m$, and $v_n$ are stored in the permanent data store, and are partitioned and indexed by sub-problems. Coefficients $b$ and $c$, parameters $\rho$, and other configurations are sent to all servers during task initialization.

Each worker obtains a list of sub-problem IDs from the orchestrator, then retrieves the corresponding coefficient partitions from the data store, and solves these sub-problems to update $X$. Each worker keeps its partitions of coefficients and intermediate result $X_k$ in memory, so as to reuse these values in subsequent iterations. Instead of sending $X_k$ to the aggregator, each worker aggregates its partition solution locally. The local aggregation results are partial left-hand-side values of constraints. Workers report them to the orchestrator.

The orchestrator keeps track of each worker's sub-problem partitions, and whether a partition has been solved in current iteration. It assigns unsolved partitions to available workers if either these partitions have not been assigned, or the assignee worker fails health check.

Once all sub-problems are solved in an iteration, the orchestrator sends partially aggregated constraint left-hand-sides to the aggregator. The aggregator is a specialized worker that knows how to update $\xi$, $\lambda$, and $\mu$. 

The system tolerates failures on workers and the aggregator, by health-checking and maintaining redundancy on servers. If a worker fails, the orchestrator will assign its job to another worker, which can retrieve coefficients and $X_k$ from the permanent data store. The aggregator is stateless, so if it fails, the aggregation task can just be sent to its backup. 

\section{Results on Synthetic Data} \label{ExperimentsSection}
In this section, we set up two classes of non-linear nonseperable problems.
We aim
to answer the following questions: (1) how can our approach approximate the
optimal solution in small assignment problems? (2) how does our post-processing of
sampling from simplex $x_i$ affect the solution of integer programming when we vary
the problem scales? (3) The scalability of our approach with 1 billion decision variables.

\subsection{Objective Functions}
If the objective function $f$ is separable by item, for example linear, we can set $g$ to $f$. The problems could
be more challenging when we have non-separable (by item) cases.
We specify two non-separable objective functions, and give their corresponding choices of $\Omega_1$. Their convexity can be proven by taking the derivatives and observe the function values are always positive.

\subsubsection{Quadratic objective}\label{QuadraticObj}

\begin{flalign}
\min_{X} & \displaystyle \quad \frac{1}{2}\sum_j(\omega_j^Tx_j+\alpha_j)^2 \label{ObjectiveQP} \\
s.t. & \displaystyle \quad X^T u_m \leq b_m, \, \forall m \in [M]  ,\quad X^T v_n = c_{n}, \, \forall n \in [N],   \\
& \displaystyle \quad \sum_j x_{ij} \leq 1, \, \forall i \in [I] , \quad 0 \leq x_{i,j} \leq 1,\, \forall i \in [I], \forall j \in [J] 
\end{flalign}

Let convex function $\Omega_1$ be
\begin{equation}
\begin{split}
\Omega_1 &= g(X) - f(X)\\
& =\sum_j \frac{N}{2}\omega_j^T\operatorname{diag}(x_j \circ x_j)\omega_j -\frac{1}{2}\sum_j(\omega_j^T x_j + \alpha_j)^2
\end{split}
\end{equation}

\subsubsection{Logrithmic objective}\label{LogObj}
\begin{flalign}
\min_{X} & \displaystyle \quad -\sum_j\operatorname{ln}(\omega_j^T x_j+a_j) \label{ObjectiveLog} \\
s.t. & \displaystyle \quad X^T u_m \leq b_m, \quad \forall m \in [M] \label{InequalityLog} \\
& \displaystyle \quad X^T v_n = c_{n}, \quad \forall n \in [N]  \label{EqualityKL} \\
& \displaystyle \quad \sum_j x_{ij} \leq 1, \quad \forall i \in [I] \label{LocalConstraintLog}\\
& \displaystyle \quad 0 \leq x_{i,j} \leq 1 \quad \forall i \in [I], \forall j \in [J] \label{IntegerConstraintLog}
\end{flalign}

Let convex function $\Omega_1$ be
\begin{equation}
\begin{split}
\Omega_1 &= g(X) - f(X)\\
& =\sum_j \frac{N}{2\alpha_j^2}\omega_j^T\operatorname{diag}(x_j \circ x_j)\omega_j +\sum_j\operatorname{ln}(\omega_j^T x_j+a_j)
\end{split}
\end{equation}

\subsection{Coefficients and Parameters}\label{SectionCoef}
If not otherwise stated, coefficients and parameters used in the subsequent experiments are generated as follows. For both (\ref{QuadraticObj}) and (\ref{LogObj}), each element of coefficient vectors $\omega_j$ and $v_n$ are drawn from uniform distribution $[0,1]$. Each element of vectors $u_m$ is drawn from uniform distribution $[-1, 0]$. Each element of vector $b_m$ is set to be $-0.3\frac{I}{J}$. Each element of vector $c_n$ is set to be $0.3\frac{I}{J}$. In (\ref{QuadraticObj}), each element of vector $\alpha_j$ is set to be $10^{-4}I$, and $\rho$ is set to $10^{-3}$. In (\ref{LogObj}), each element of vector $\alpha_j$ is set to be $10^{-1}I$, and $\rho$ is set to $10^{-5}$. $\beta$ is set to $\frac{\rho}{2}I(M + N)$. $I$, $J$, $M$, $N$, and number of iterations vary in different experiments.

\subsection{Algorithm Correctness}
\subsubsection{Evenly distributed coefficients}\label{SectionEvenExperiment}
We compared the objective values of the solution obtained by our implementation and that of the open-sourced solver IPOPT \cite{wachter2006implementation}. IPOPT is a commonly used non-linear constrained optimization solver. We use it as a reference, and argue that an optimal solution should be close to an IPOPT solution. We use absolute percentage difference (APD) to quantify the difference between the two solutions. We use $\text{sol}$ to represent the objective value of ADMM solution, and $\text{sol}*$ to represent the objective value of IPOPT.
\begin{equation}
\text{APD} := |\frac{\text{sol}-\text{sol}^*}{\text{sol}^*}|
\end{equation}

To quantify the violation of an inequality constraint, we use the left-hand-side subtract the right-hand-side of the inequality, and set it to zero if this difference is negative. For example, the $j$-th constraint for a specific $m$ of (\ref{Inequality}) is $x_j^Tu_m \leq b_{mj}$, and its APD is
\begin{equation}
\text{APD} := |\frac{\max({x_j^Tu_m-b_{mj}, 0})}{b_{mj}}|
\end{equation}

To quantify the violation of an equality constraint, we use the relative difference between the left-hand-side and right-hand-side. For example, the $j$-th constraint for a specific $n$ of (\ref{Equality}) is $x_j^Tv_n=c_{nj}$, and the APD is
\begin{equation}
\text{APD} := |\frac{x_j^Tv_n-c_{nj}}{c_{nj}}|
\end{equation}

With the solution to one problem, we take the mean of the APD for all its inequality constraints to get the inequality mean APD (MAPD), and the mean of the APD for all its equality constraints to get the equality MAPD.

The results of the correctness metrics are summarized in Table \ref{TableQuadraticCorrectness} and Table \ref{TableLogCorrectness}. To solve the problems with IPOPT, which runs on single machine, we set $I=3000$. We tested on various values of $M$, $N$, and $J$. To solve the quadratic objective problems, we let ADMM run 2500 iterations, and found the max objective APD is less than 0.04, the max inequality MAPD is less than 0.001, and the max equality MAPD is less than 0.0015. To solve the logarithmic objective problems, we let ADMM run 1000 iterations, and found the max objective APD is less than 0.05, the max inequality MAPD is 0, and the max equality MAPD is less than 0.0008. These difference metrics indicate the ADMM solutions are close to optimal with very small constraint violation.

\begin{table}
\caption{Results of quadratic objective problems}
\label{TableQuadraticCorrectness}
\centering
\begin{tabular}{cccccc}
\toprule
$M$ & $N$ & $J$ & Obj. APD & Ineq. MAPD & Eq. MAPD\\
\midrule
3 & 2 & 10 & 0.0032 & 0.0006 & 0.0006 \\
5 & 2 & 10 & 0.0069 & 0.0008 & 0.0011 \\
10 & 2 & 10 & 0.0250 & 0.0006 & 0.0014 \\
3 & 5 & 10 & 0.0385 & 0.0009 & 0.0010 \\
3 & 10 & 10 & 0.0256 & 0.0009 & 0.0010 \\
3 & 2 & 15 & 0.0351 & 0.0009 & 0.0006 \\
\bottomrule
\end{tabular}
\end{table}

\begin{table}
\caption{Results of logarithmic objective problems}
\label{TableLogCorrectness}
\centering
\begin{tabular}{cccccc}
\toprule
$M$ & $N$ & $J$ & Obj. APD & Ineq. MAPD & Eq. MAPD\\
\midrule
3 & 2 & 10 & 0.0421 & 0.0 & 0.0004 \\
5 & 2 & 10 & 0.0420 & 0.0 & 0.0004 \\
10 & 2 & 10 & 0.0414 & 0.0 & 0.0004 \\
3 & 5 & 10 & 0.0379 & 0.0 & 0.0005 \\
3 & 10 & 10 & 0.0353 & 0.0 & 0.0006 \\
3 & 2 & 15 & 0.0326 & 0.0 & 0.0008 \\
\bottomrule
\end{tabular}
\end{table}

\subsubsection{Unevenly distributed coefficients}
When different sub-problems have different coefficient distributions, solving the optimization problems become more difficult. To test the quality of our ADMM solution in such scenarios, we mix 1500 items which have coefficient values drawn from the distributions described in \ref{SectionCoef}, with another 1500 items which have coefficients scaled 10 times, i.e. $\omega_j$ and $v_n$ are drawn from uniform distribution $[0,10]$, and $u_m$ is drawn from uniform distribution $[-10, 0]$. Other coefficients and parameters are the same as in \ref{SectionEvenExperiment}.

The results are summarized in Table \ref{TableQuadraticCorrectnessUneven} and Table \ref{TableLogCorrectnessUneven}. The quadratic objective problems have max objective APD less than 0.05, max inequality MAPD less than 0.002, and max equality MAPD less than 0.0003. The logarithmic objective problems have max objective APD less than 0.09, max inequality MAPD as 0, and max equality MAPD less than 0.0009. These difference metrics indicate that when items have drastically (10x) different coefficient distributions, constraint violation remains at the same level, while solution optimality decreased, at least in the logarithmic problem.

\begin{table}
\caption{Results of quadratic objective problems with unevenly distributed coefficients}
\label{TableQuadraticCorrectnessUneven}
\centering
\begin{tabular}{cccccc}
\toprule
$M$ & $N$ & $J$ & Obj. APD & Ineq. MAPD & Eq. MAPD\\
\midrule
3 & 2 & 10 & 0.0446 & 0.0002 & 0.0003 \\
5 & 2 & 10 & 0.0039 & 0.0002 & 0.0002 \\
10 & 2 & 10 & 0.010 & 0.0001 & 0.0002 \\
3 & 5 & 10 & 0.0004 & 0.0002 & 0.0003 \\
3 & 10 & 10 & 0.0114 & 0.0002 & 0.0003 \\
3 & 2 & 15 & 0.0183 & 0.0001 & 0.0002 \\
\bottomrule
\end{tabular}
\end{table}

\begin{table}
\caption{Results of logarithmic objective problems with unevenly distributed coefficients}
\label{TableLogCorrectnessUneven}
\centering
\begin{tabular}{cccccc}
\toprule
$M$ & $N$ & $J$ & Obj. APD & Ineq. MAPD & Eq. MAPD\\
\midrule
3 & 2 & 10 & 0.0857 & 0.0 & 0.0009 \\
5 & 2 & 10 & 0.0859 & 0.0 & 0.0009 \\
10 & 2 & 10 & 0.086 & 0.0 & 0.0006 \\
3 & 5 & 10 & 0.0627 & 0.0 & 0.0008 \\
3 & 10 & 10 & 0.0343 & 0.0 & 0.0006 \\
3 & 2 & 15 & 0.0541 & 0.0 & 0.0010 \\
\bottomrule
\end{tabular}
\end{table}

\subsection{Effects of Integer Relaxation}
We evaluate constraint MAPD as described in \ref{SectionEvenExperiment}. We evaluate the solution objective by comparing it to the continuous solution objective, which can be viewed as a lower bound. If we represent the continuous solution objective as $\text{lb}$, the objective APD is
\begin{equation}
\text{APD} := |\frac{\text{sol}-\text{lb}}{\text{lb}}|
\end{equation}

With coefficients and parameters set the same as in \ref{SectionEvenExperiment}, Table \ref{TableQuadraticCorrectnessInteger}
 and Table \ref{TableLogCorrectnessInteger} summarize the solution quality. The quadratic objective problems have max objective APD less than 0.025, max inequality MAPD less than 0.015, and max equality MAPD less than 0.015. The logarithmic objective problems have max objective APD less than 0.002, max inequality MAPD as 0, and max equality MAPD less than 0.05. These difference metrics indicate that with tens of thousands of decision variables (30,000-45,000), integer post-processing generates objectives close to their relaxation lower bounds, and that constraint violation increases, though still at low level.
 
\begin{table}
\caption{Results of quadratic objective problems with integer post-processing}
\label{TableQuadraticCorrectnessInteger}
\centering
\begin{tabular}{cccccc}
\toprule
$M$ & $N$ & $J$ & Obj. APD & Ineq. MAPD & Eq. MAPD\\
\midrule
3 & 2 & 10 & 0.0080 & 0.0030 & 0.0046 \\
5 & 2 & 10 & 0.0119 & 0.0087 & 0.0084 \\
10 & 2 & 10 & 0.0113 & 0.0131 & 0.0136 \\
3 & 5 & 10 & 0.0053 & 0.0064 & 0.0111 \\
3 & 10 & 10 & 0.0118 & 0.0065 & 0.0086 \\
3 & 2 & 15 & 0.0211 & 0.0049 & 0.0128 \\
\bottomrule
\end{tabular}
\end{table}

\begin{table}
\caption{Results of logarithmic objective problems with integer post-processing}
\label{TableLogCorrectnessInteger}
\centering
\begin{tabular}{cccccc}
\toprule
$M$ & $N$ & $J$ & Obj. APD & Ineq. MAPD & Eq. MAPD\\
\midrule
3 & 2 & 10 & 0.0001 & 0.0 & 0.0386 \\
5 & 2 & 10 & 0.0 & 0.0 & 0.0429 \\
10 & 2 & 10 & 0.0 & 0.0 & 0.0423 \\
3 & 5 & 10 & 0.0001 & 0.0 & 0.0321 \\
3 & 10 & 10 & 0.0002 & 0.0 & 0.0423 \\
3 & 2 & 15 & 0.0002 & 0.0 & 0.0345 \\
\bottomrule
\end{tabular}
\end{table}

\subsection{Scalability and Convergence}
We then scale up the problem sizes to $I=10^8$. We set $J=10$, $M=3$, and $N=2$. Thus the number of decision variables is $1$ billion. This scale corresponds to the scenario of making one-out-of-ten choices for 100 million items, which is not uncommon for internet companies.

\begin{figure}
\centering
\includegraphics[width=0.6\linewidth]{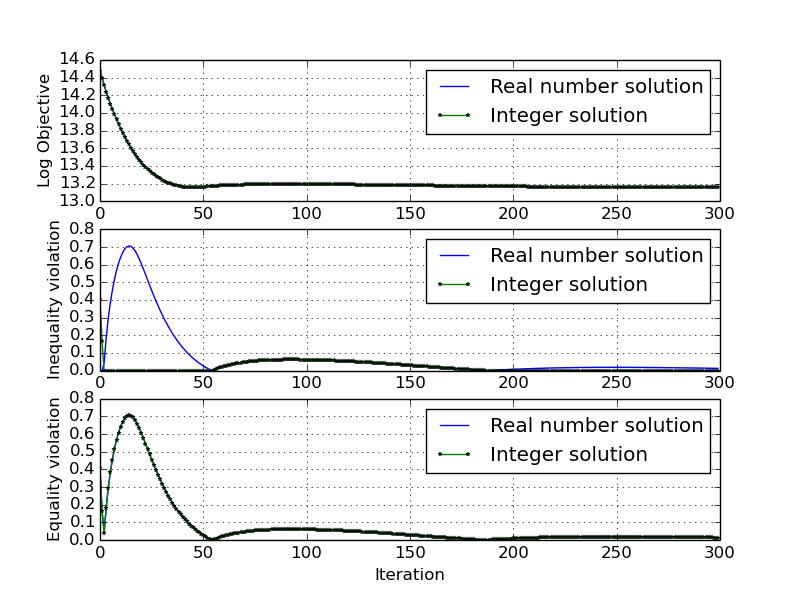}
\caption{Quadratic Objective Problem Convergence}
\label{convergencePlotQP}
\end{figure}

Figure \ref{convergencePlotQP} shows objective (log10 scale), inequality MAPD, and equality MAPD versus ADMM iterations, for the quadratic objective problem. At 2000 iterations, inequality MAPD is 0.00026 and equality MAPD 0.00036. After integer post-processing by sampling probability simplex, inequality MAPD is still 0.00026 and equality MAPD is still 0.00036. The objective curve has converged, and the APD between continuous and integer solution is $3.6 \times 10^{-6}$. We only show the result of the first 300 iterations as the remaining curves look flat.

\begin{figure}
\centering
\includegraphics[width=0.6\linewidth]{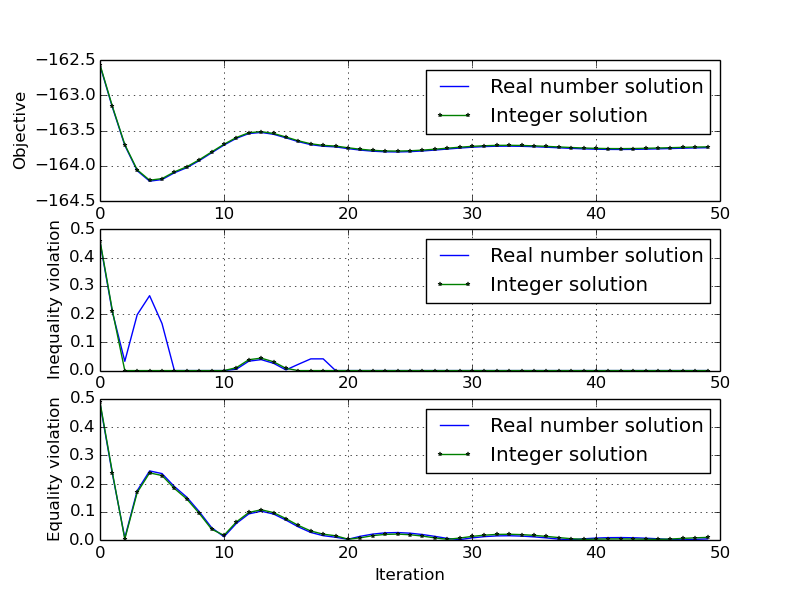}
\caption{Logarithmic Objective Problem Convergence}
\label{convergencePlotLog}
\end{figure}

Figure \ref{convergencePlotLog} shows the same metrics for the logarithmic objective problem. At 100 iterations, inequality MAPD is zero, equality MAPD is $10^{-5}$. After integer post-processing by sampling probability simplex, inequality MAPD is still zero, and equality MAPD is still $10^{-5}$. The objective curve has converged, and the APD between continuous and integer solution is $7.2 \times 10^{-5}$. We only show the result of the first 50 iterations as the remaining curves look flat.

We implemented the algorithm using the programming language Python. When the software runs on 2000 CPUs, the average time consumed per iteration is 100 seconds. This includes all steps in Section \ref{ADMMMethod}. The implementation features "horizontal scalability", so can further scale out.

\section{Real-world Applications}\label{real-world}
Solving assignment problems at scale is crucial for many real world businesses. We developed our distributed ADMM solver based on the AntOpt platform~\cite{zhou2021antopt}, and have made it in production at Alipay\footnote{\url{https://en.wikipedia.org/wiki/Alipay}} since mid 2021. It has been used to power crucial decisions such as allocation and reallocation of several hundred million users in loans, capital management of financial institutes, portfolio management, and marketing, etc. Our billion-scale ADMM solver has supported running of both daily and on-demand tasks in our business.

\subsection{Financial Product Marketing}
In this section, we describe a concrete application in marketing. There are $J$
equity funds that are managed by different owners. We aim to assign $J$ funds
to $I$ users while each user can be exposed with at most one equity fund. The
owners themself require that the population of the users to be assigned should
satisfy specific demographics. That is, we are given $K$ groups of users
$\{1,...,k,...,K\}$ that each group defines an exclusive set of users. For example,
group $k=1$ denotes a set of users who are sensitive to the risk of equity funds and contribute to high volume and open interest, and group $k=2$ denotes a set of users who are sensitive to the risk but contribute low volume and open interest, and so
on so forth. 
In reality, each owner $j$ would specify a desired population of users denoted as 
the distribution $q_{j,k}$. We are required to make the assignment to satisfy each owner's desired population. Typically we use population stability index (PSI)
\footnote{PSI is an important metric to identify a shift in 
population for retail credit scorecards~\cite{karakoulas2004empirical}.} to
measure the matching of the population, i.e., $\sum_{k=1}^{K} (p_{j,k} - q_{j,k}) \ln \frac{p_{j,k}}{q_{j,k}}, \forall j$ where we denote $p_{j,k}$ as the population
as we do the assignment for each owner $j$. We aim to make the assignment
to minimize the PSI of each owner. Besides, each owner has an inventory of supply
that is prepared in advance. Our assignment should help meet the supply of each
owner. As such, we have the following assignment problem:
\begin{equation}
\begin{aligned}
\min & \sum_j \sum_{k} (\frac{\sum_{i\in I_k} c_{ij} x_{ij}}{b_j} - q_{j,k}) \ln \frac{\frac{\sum_{i\in I_k} c_{ij} x_{ij}}{b_j}}{q_{j,k}} \\
&s.t. \sum_i c_{ij} x_{ij} = b_j, \,\,\, \forall j \quad \text{and} 
\quad \sum_j x_{ij} \leq 1, \,\,\, \forall i,
\end{aligned}
\end{equation}
where we denote $c_{ij}$ as the probability of purchase of fund $j$ by user $i$,
$b_j$ as the inventory of fund $j$, and $I_k$ as the set of users in the
$k$-th group. In practice, each day we assign around 1.1 hundred of million users
to tens of funds to meet the demand of fund owners via notification
in Alipay app. Note that the objective is nonseperable due to the $\log$ of a sum.

\subsection{Results}
\subsubsection{Numeric Results}
\begin{figure}
\centering
\includegraphics[width=0.8\linewidth]{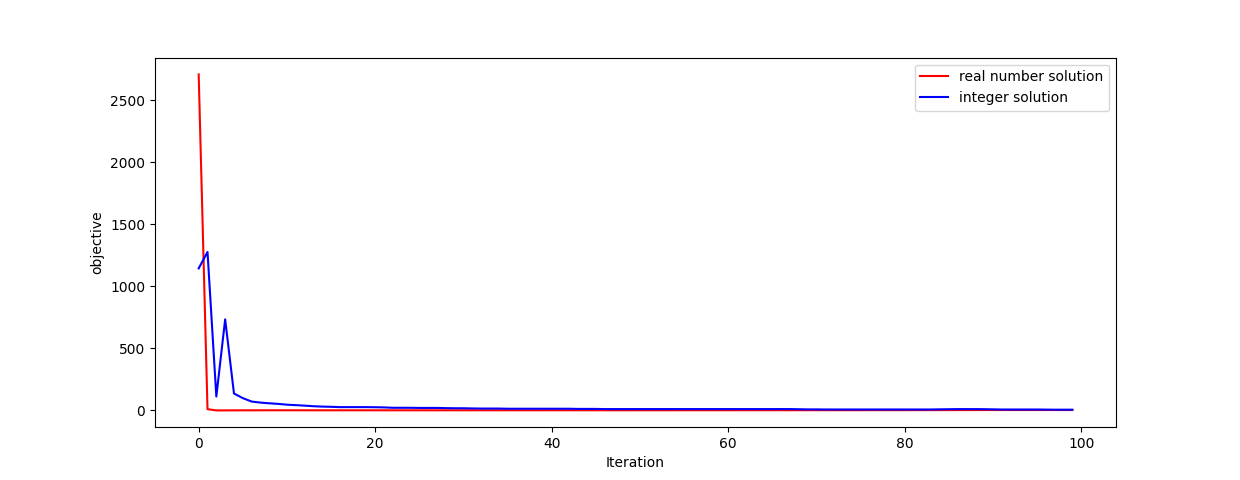} \\
\includegraphics[width=0.8\linewidth]{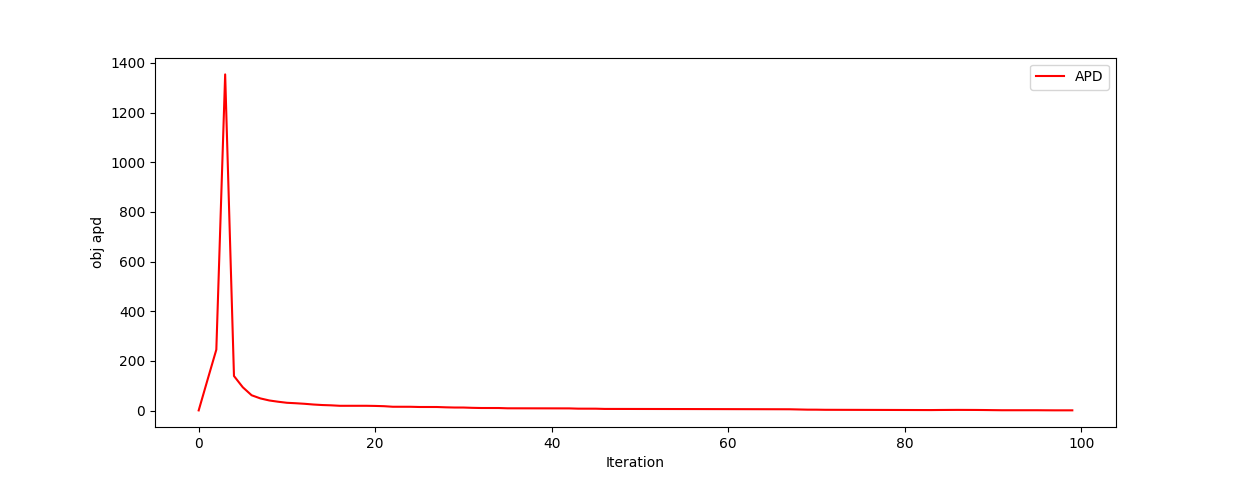} \\
\includegraphics[width=0.8\linewidth]{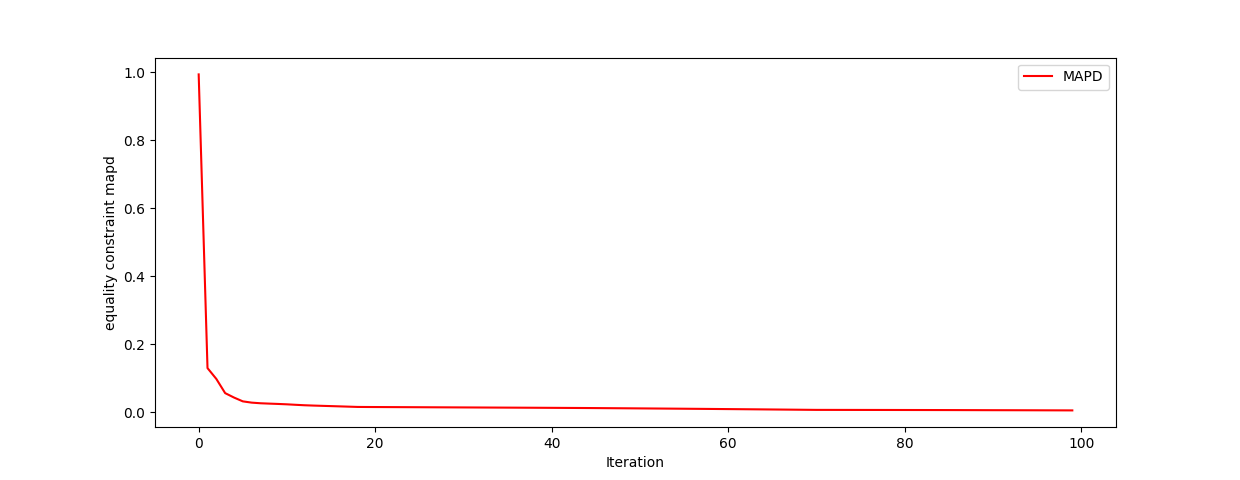}
\caption{Convergence of users-fund assignment problem.}
\label{fig:objective}
\end{figure}

Following the case of financial product marketing, we have nearly
1.1 hundred of million users and 21 equity fund owners each day.
We show the results of our algorithm for solving this real-world
large scale problem in Figure~\ref{fig:objective}. We first
show the objectives w.r.t. real number solution and integer solution
and their APD as our algorithm converges. It shows that the objective gap between
integer solution and real number solution is large at the first few round
of iterations but converges to 0.00231 as the real number solutions approximate
to either 0 or 1. We also show the MAPD of equality constraint that the violation
being consistently optimized up to 0.00081. 
\begin{figure}
\centering
\includegraphics[width=0.6\linewidth]{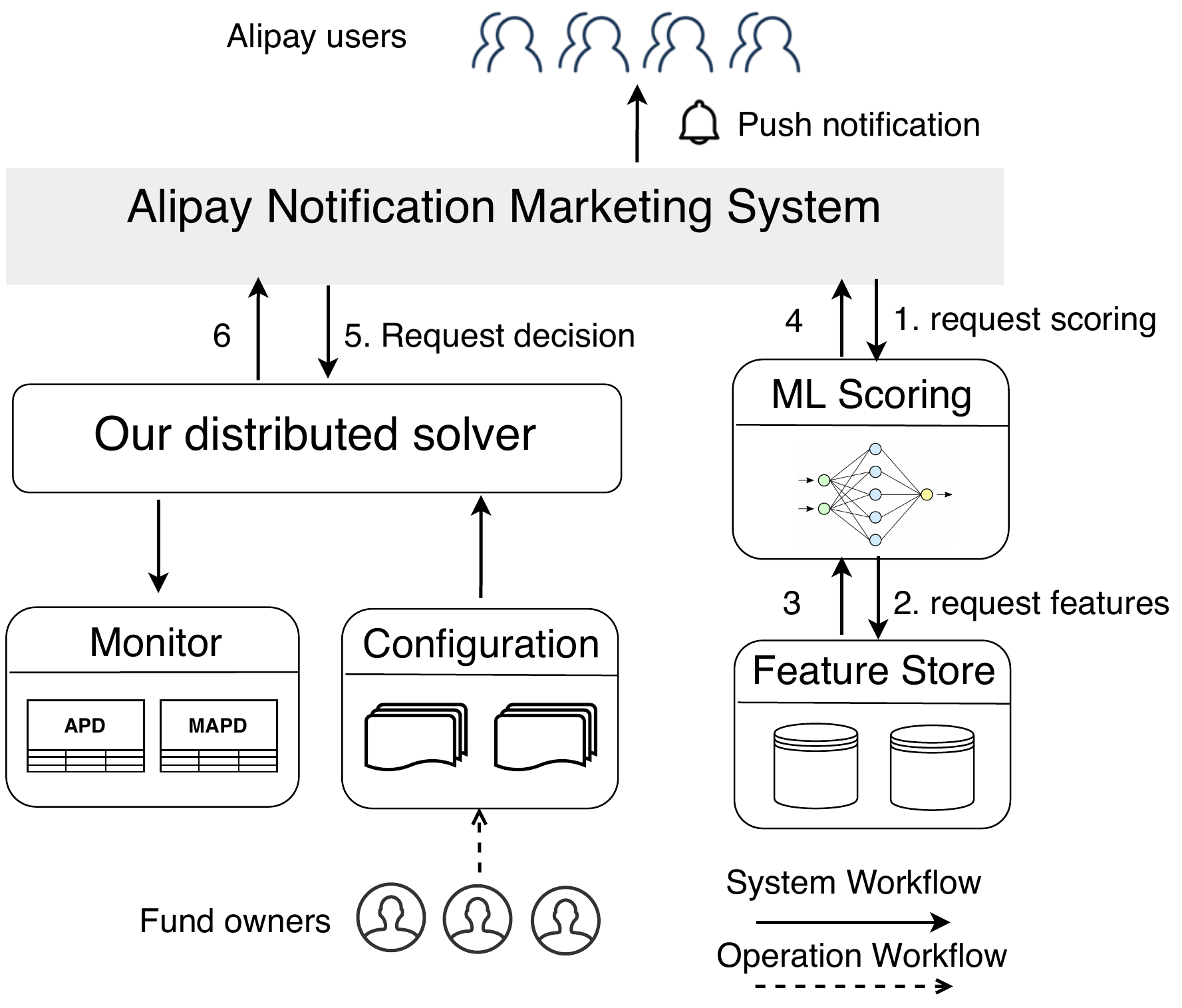}
\caption{Workflow of our billion scale assignment solver.}
\label{fig:workflow}
\end{figure}

\subsubsection{Deployment and Online Results}
We first describe the deployment of our distributed solver, then
present the online results. We demonstrate the overall workflow
in Figure~\ref{fig:workflow}. We have fund owners and Alipay users.
We allow fund owners to config parameters such as inventory of supply and desired
population of users. The Alipay notification marketing system maintains the
overall workflow. Note that the overall decision making is data-driven based,
i.e., it relies on machine learning components to predict the scores
such as the probability of purchase $c_{i,j}$. The system first requests
the ``ML scoring'' component to retrieve the scores (i.e., coefficients of our assignment problem) for final decision making. The ``ML scoring'' component requests ``Feature store'' to access the
necessary features. After returning the scores back to ``Notification system'',
the system sends such coefficients to our distributed solver (see
details in Figure~\ref{SystemGraph}). After solving the assignment problem, the solver
returns the decisions and finally completes the notification to Alipay users.
We further have monitors to record the quality of our solutions.

The fund owners prepare the inventory and configure the numbers each day.
They may also periodically adjust the desired population $q_{j,k}$. Given
those parameters, we solve the problem each day. Note that the lower objective value the better, and the lower mean equality violation (MAPD) the better. We report the average relative
improvements w.r.t. the quality of our solution compared with a heuristic
approach in 7 days in Table~\ref{tb:OnlineResult}.

\begin{table}
\caption{Relative improvement in online results.}
    \centering
    {%
    \begin{tabular}{lrr}
        \toprule
        Metrics & Our solver \\
        \midrule
        Objective  & \textbf{14.20\%} improvement \\
        Mean Equality Violation  & \textbf{+7.20\%} improvement \\
        \bottomrule
    \end{tabular}}
    
    \label{tb:OnlineResult}
\end{table}

\section{Conclusions} \label{conclusion}

We introduced the formulation for a class of assignment problems. The formulation supports nonlinear objective, convex per-item constraints, and feature-owner cross constraints. It is more general than GAP, and covers many real-world business problems. In the era of big data, there are scenarios when solving such problems with billions of decision variables is essential to business. To achieve such scale, we adapted the Bregman ADMM framework to our problem and obtained a distributed algorithm. In experiments, we specified the full Bregman ADMM construction for two example problems, with quadratic and logarithmic objectives, respectively. Using synthetic data, we showed the correctness of our implementation, and scaled the solver to problems of one billion decision variables. We deploy our distributed solver to production and show its application to business decision making in an internet scale.

\bibliographystyle{ACM-Reference-Format}
\balance
\bibliography{main_arvix}

\end{document}